\documentclass[runningheads]{llncs}
\usepackage{array}
\usepackage{tabularx}
\usepackage{subcaption}  
\usepackage{amsmath}
\usepackage{amssymb}
\usepackage{float}      
\usepackage{adjustbox}  
\usepackage{multirow}   
\usepackage[colorlinks=true,citecolor=blue,linkcolor=blue]{hyperref}

\usepackage[T1]{fontenc}
\usepackage{graphicx,verbatim}
\begin{document}
%
\title{Multiscale Structure-Guided Latent Diffusion for Multimodal MRI Translation}
\titlerunning{Multiscale Structure-Guided MRI Translation}
%

\author{ Jianqiang Lin\inst{1,2}\textsuperscript{†} \and Zhiqiang Shen\inst{1,2}\textsuperscript{†} \and Peng Cao\inst{1,2,3}\textsuperscript{} \and Jinzhu Yang\inst{1,2,3} \and Osmar R. Zaiane\inst{4} \and Xiaoli Liu\inst{5} } \institute{ Computer Science and Engineering, Northeastern University, Shenyang, China \and Key Laboratory of Intelligent Computing in Medical Image of Ministry of Education, Northeastern University, Shenyang, China \and National Frontiers Science Center for Industrial Intelligence and Systems Optimization, Shenyang, China \\ \email{caopeng@cse.neu.edu.cn} \and Amii, University of Alberta, Edmonton, Alberta, Canada \and AiShiWeiLai AI Research, Beijing, China } 
  
\maketitle              

\begin{abstract}
Although diffusion models have achieved remarkable progress in multi-modal magnetic resonance imaging (MRI) translation tasks, existing methods still tend to suffer from anatomical inconsistencies or degraded texture details when handling arbitrary missing-modality scenarios. To address these issues, we propose a latent diffusion-based multi-modal MRI translation framework, termed \textbf{MSG-LDM}. By leveraging the available modalities, the proposed method infers complete structural information, which preserves
reliable boundary details. Specifically, we introduce a style--structure disentanglement mechanism in the latent space, which explicitly separates modality-specific style features from shared structural representations, and jointly models low-frequency anatomical layouts and high-frequency boundary details in a multi-scale feature space. 
During the structure disentanglement stage, high-frequency structural information is explicitly incorporated to enhance feature representations, guiding the model to focus on fine-grained structural cues while learning modality-invariant low-frequency anatomical representations. 
Furthermore, to reduce interference from modality-specific styles and improve the stability of structure representations, we design a style consistency loss and a structure-aware loss. Extensive experiments on the BraTS2020 and WMH datasets demonstrate that the proposed method outperforms existing MRI synthesis approaches, particularly in reconstructing complete structures. The source code is publicly available at \url{https://github.com/ziyi-start/MSG-LDM}.
\end{abstract}
\begin{figure}[H]  
	\centering
	\includegraphics[width=0.9\textwidth]{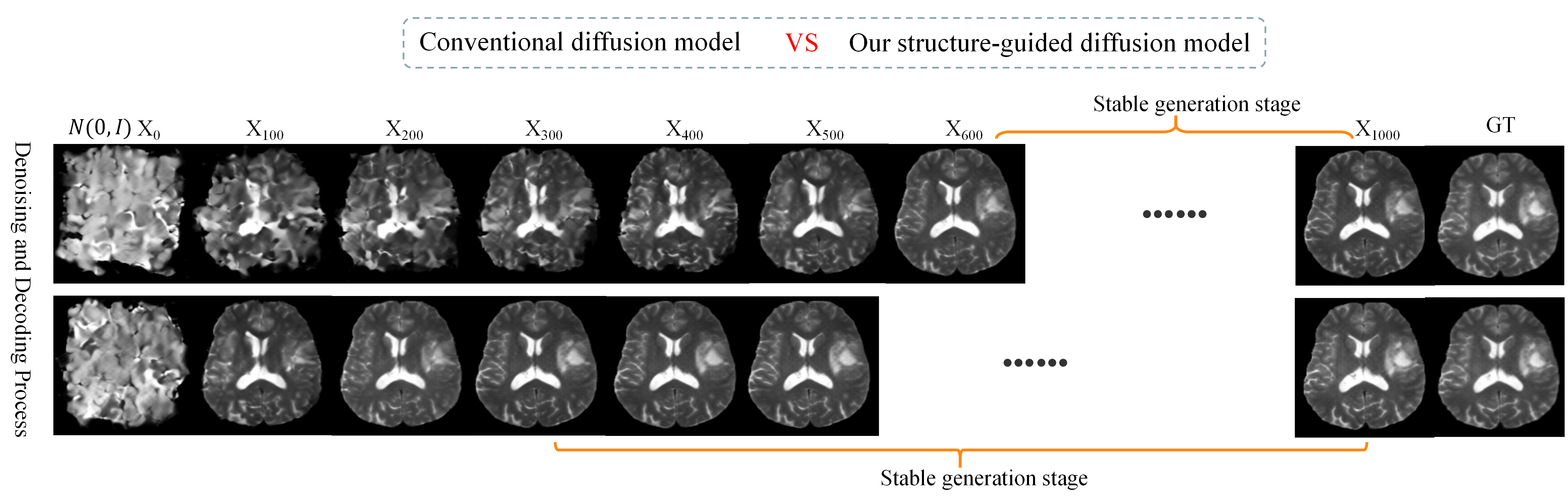}
	\caption{Denoising comparison between the conventional diffusion model and our structure-guided diffusion model at identical diffusion timesteps. Numerical subscripts indicate the denoising timestep.}
	\label{fig:denoise}
\end{figure}

\section{Introduction}

Magnetic Resonance Imaging (MRI) is a non-invasive imaging modality with high soft-tissue contrast and plays a critical role in brain disease diagnosis, grading, and treatment monitoring. Multi-modal MRI acquisitions, including T1-weighted, T2-weighted, contrast-enhanced T1-weighted, and FLAIR sequences, provide complementary anatomical and pathological information and are widely used for brain tumor segmentation and lesion analysis. However, due to long acquisition times, limited patient tolerance, and equipment or cost constraints, complete multi-modal MRI data are often unavailable in clinical practice, resulting in missing modalities and significantly degrading the performance of multi-modal analysis algorithms and clinical reliability~\cite{ref_article1,ref_article2,ref_article3}.

Recently, diffusion models have demonstrated strong performance in image generation and have been increasingly adopted for medical image synthesis. By learning the reverse data distribution through progressive noise injection and iterative denoising, diffusion models exhibit superior training stability and structural detail preservation~\cite{ref_article9,ref_article10}. Existing studies have shown that diffusion-based approaches outperform GANs in structural fidelity and visual quality for cross-modal MRI synthesis~\cite{ref_article11,ref_article12,ref_article20,ref_article13,ref_article14,ref_article15}. 

Despite progress, diffusion-based methods still struggle in multi-modal image-to-image translation under arbitrary missing-modality scenarios, where anatomical structures may distort, high-frequency details degrade, and structural information entangles with modality-specific styles, limiting synthesis fidelity and consistency.  
Traditional diffusion models lack structural awareness, leading to inefficient and unstable reconstruction. As shown in Fig.~1, introducing a structural prior in our method improves structural stability and accelerates generation.  
To address these challenges, we propose \textbf{MSG-LDM}, a latent diffusion-based framework for multi-modal MRI translation. The main contributions are summarized as follows:
\begin{enumerate}
\item \textbf{Structure-Guided Latent Diffusion.}
	We demonstrate that diffusion models are intrinsically insensitive to structural information in medical images. By explicitly incorporating structural priors, the proposed framework substantially accelerates the generation process while preserving anatomical fidelity.

	\item \textbf{Multi-modality Multi-scale Structural Representation Learning.} 
	We design a structural encoder with a high-frequency injection block, multi-modal structural feature fusion, and multi-scale structure feature enhancement to capture the modality-invariant structural representations, involving low-frequency structural context and high-frequency boundary details. Meanwhile, we introduce a \textbf{style consistency loss} and a \textbf{structure-aware loss} to regularize the \textbf{multi-scale structural representations}, thereby enhancing structural integrity, suppressing modality-specific style interference, and improving cross-modality consistency.

	\item \textbf{Enhanced cross-modal MRI synthesis Performance.}  
	Extensive experiments on the BraTS2020 and WMH datasets demonstrate that MSG-LDM consistently outperforms representative state-of-the-art methods in anatomical preservation and fine-detail reconstruction quality, validating its effectiveness and robustness.
\end{enumerate}

\section{Method}
\subsection{Overview of the Framework}
As shown in Fig.~2, we propose a \textbf{multi-scale structure-guided Latent Diffusion Model (MSG-LDM)}, which performs the diffusion process in the VAE latent space~\cite{ref_article16,ref_article29,ref_article23}.
The image representations are then disentangled into structure feature and style feature.  
The core aim of our method is to perform multi-modal, multi-scale structural feature disentanglement from inherently complex and highly variable medical images. This disentanglement reduces modality-specific style interference while extracting complete structural information, including low-frequency global anatomical structures and high-frequency edge and texture details. These structural priors guide the diffusion process, enabling effective synthesis of missing modalities.

\begin{figure}[htbp]
	\centering
	\includegraphics[width=0.9\textwidth]{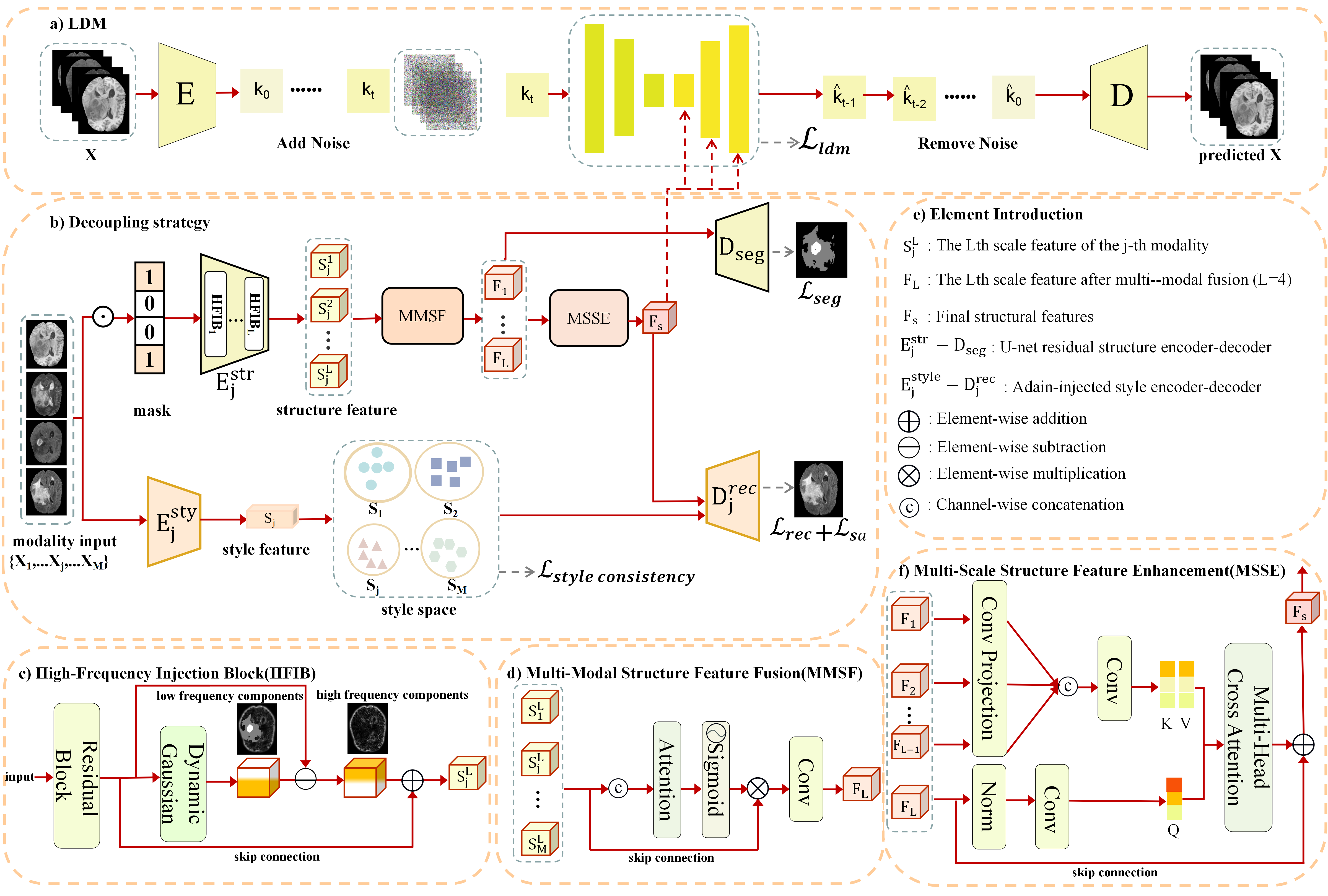}
	\caption{Overview of the proposed MSG-LDM for multi-modal MRI synthesis.}
	\label{fig1}
\end{figure}

Given multi-modal inputs $\{X_j\}_{j=1}^{M}$, we first apply a partial masking strategy to simulate missing modality scenarios. 
Each modality is equipped with a distinct structural encoder with a High-Frequency Injection Block (HFIB) $E_j^{\mathrm{str}}$, a style encoder $E_j^{\mathrm{sty}}$, and a reconstruction decoder $D_j^{\mathrm{rec}}$, while all modalities share a segmentation decoder $D_{\mathrm{seg}}$ ($j=1,\dots,M$) for ensuring the structure feature being modality invariant.  
The multi-scale structural features extracted by $E_j^{\mathrm{str}}$ are first fused across modalities at each scale via the Multi-Modal Structure Feature Fusion (MMSF) module, and then enhanced through the Multi-Scale Structure Feature Enhancement (MSSE) module to produce a unified structural representation $F_s$ (see Section~2.2 for details). This representation guides the diffusion-based generation process.  
Conditioned on the unified structural representation $F_s$, the LDM is trained using the standard diffusion loss:
\begin{equation}
	\mathcal{L}_{\text{LDM}} =
	\mathbb{E}_{z_0,\, F_s,\, t,\, \epsilon}
	\left\|
	\epsilon - \epsilon_\theta(z_t,\, t \mid F_s)
	\right\|_2^2.
\end{equation}

\subsection{Structure feature learning}
As shown in Fig.~2(b), the structure encoder $E_j^{\mathrm{str}}$ for the $j$-th modality extracts multi-scale features $\{ S_j^{(l)} \}_{l=1}^{L}$, where $L=4$ indicates the number of scales. At each scale, cross-modal features are adaptively fused via \textbf{Multi-Modal Structural Feature Fusion (MMSF)}, which combines features from all modalities using learnable attention weights to emphasize informative structures while suppressing irrelevant modality-specific variations. The fused features $\{ F_l \}_{l=1}^{L}$ are then fed into the segmentation decoder $D_{\mathrm{seg}}$ for prediction and further refined by \textbf{Multi-Scale Structure Feature Enhancement (MSSE)} to obtain the unified structural representation
$F_s$, capturing both global anatomical layouts and fine-grained structural details. 
Additionally, we introduce a \textbf{structure-aware loss} to jointly model low- and high-frequency components. Meanwhile, the style encoder extracts modality-specific style features $S_j = E_j^{\mathrm{sty}}(X_j)$, with a \textbf{style consistency} imposed to suppress modality-specific style interference, providing effective supervision for the feature disentanglement process and enabling the learning of robust and structurally consistent representations.

\subsubsection{High-Frequency Injection Block (HFIB)}

HFIB is designed to enhance high-frequency structural information at multiple scales within the structure encoder, enabling the network to preserve low-frequency global anatomy while emphasizing high-frequency edges and texture details~\cite{ref_article21,ref_article17,ref_article18,ref_article19}, as illustrated in Fig.~2(c). 

Given the content feature at the $l$-th scale, $C^{l}$, HFIB decomposes it via a learnable dynamic Gaussian filter:
\begin{equation}
	C_{\mathrm{high}}^{l} = C^{l} - \mathcal{G}_{\theta_l}(C^{l}), \quad
	S_j^{l} = C^{l} + C_{\mathrm{high}}^{l},
\end{equation}
where $\mathcal{G}_{\theta_l}(\cdot)$ is a learnable, input-adaptive Gaussian filtering operator at the $l$-th scale. Here, $\mathcal{G}_{\theta_l}(C^{l})$ captures the low-frequency components, while $C_{\mathrm{high}}^{l}$ represents high-frequency residuals such as edges and fine textures. By re-injecting $C_{\mathrm{high}}^{l}$ into the original feature, $S_j^{l}$ enhances structural details without altering the global anatomical layout.

\subsubsection{Multi-Modal Structural Feature Fusion (MMSF)}  
As shown in Fig.~2(d), at each scale $l$, MMSF fuses structural features from $M$ modalities $\{ S_j^{(l)} \}_{j=1}^{M}$ by computing attention weights $w_j$ via a Sigmoid gating network and combining them through a learnable convolution, such that $F_l = \mathrm{Fusion}\left( \sum_{j=1}^{M} w_j S_j^{(l)} \right)$, where $F_l$ is the fused feature and $w_j \in [0,1]$ denotes the attention for the $j$-th modality.

\subsubsection{Multi-Scale Structural Feature Enhancement (MSSE)}

As illustrated in Fig.~2(f), given the multi-scale shared structural features 
\(\{F_l\}_{l=1}^{L}\) extracted by the structure encoder and fused via MMSF, 
MSSE injects multi-scale high-frequency structural information into high-level 
representations through structure-guided cross-attention. Features from lower 
scales (\(1\) to \(L-1\)) are first projected and spatially aligned to the highest 
scale to form a structural guidance feature. The highest-scale representation 
is then enhanced via cross-attention followed by residual fusion,  
\( F_s = F_L + \alpha \, \mathrm{Attn}\Big(F_L, \sum_{l=1}^{L-1} \mathrm{Up}(\mathrm{Proj}(F_l)) \Big) \), 
where \(\mathrm{Proj}(\cdot)\) and \(\mathrm{Up}(\cdot)\) denote \(1\times1\) 
convolution and bilinear upsampling, respectively. The resulting unified 
structural representation \(F_s\) integrates low-frequency global anatomical 
layouts and high-frequency detailed structures, making it suitable for 
subsequent reconstruction and diffusion processes.

\subsection{Overview of Loss Functions}
\subsubsection{Style Consistency Loss}

To mitigate modality-specific style interference during structure disentanglement, we introduce a style consistency loss, analogous to contrastive learning. Let $S_{b,j}$ denote the style feature of modality $j$ from sample $b$, where $j \in \{1,\dots,M\}$ with $M$ being the number of modalities, and $b \in \{1,\dots,B\}$ with $B$ being the mini-batch size. All style features within the mini-batch are $L_2$-normalized and stacked as $\{\tilde{S}_p\}_{p=1}^{M \times B}$.  
Pairwise similarities are computed via a temperature-scaled dot product, $z_{pq} = \tau \tilde{S}_p^\top \tilde{S}_q$, and a binary cross-entropy objective is applied:

\[
L_\mathrm{sc} = - \frac{1}{(M \times B)^2} \sum_{p,q=1}^{M \times B} \Big[ T_{pq} \log \sigma(z_{pq}) + (1-T_{pq}) \log \sigma(-z_{pq}) \Big],
\]

where $\sigma(\cdot)$ is the sigmoid function and $\tau$ is a learnable temperature, if $T_{pq}=1$, \textit{i.e.,} the samples $p$ and $q$ belong to the same modality, it pulls together their style; if $T_{pq}=0$, it pushes apart style features across different modalities. This design encourages the style encoder to suppress modality-specific style variations while maintaining consistency across the same modality.

\subsubsection{Structure-aware Loss.} 
To effectively supervise multi-scale feature disentanglement while preserving anatomical structures and fine-grained details, we propose a structure-aware loss, which consists of two components: a reconstruction loss and a frequency-domain SSIM loss. 
With the disentangled modality-invariant structure features $F_s$ and the $j$-th modality style features $S_j$, the reconstruction decoder produces $\hat{X}_j = D_j^{\mathrm{rec}}(F_s, S_j)$, to further guide the structure feature disentanglement.
The reconstruction loss uses the $L_1$ norm to enforce overall voxel-level intensity fidelity between the generated and ground-truth images, while the frequency-domain SSIM loss encourages global structure consistency by comparing the magnitude spectra of the generated and ground-truth images after a 2D discrete cosine transform (DCT):
\[
L_\mathrm{sa} = L_\mathrm{rec} + L_\mathrm{freq}, \quad L_\mathrm{freq} = 1 - \mathrm{SSIM}\big(|\mathcal{D}(\hat{X}_j)|,\ |\mathcal{D}(X_j)|\big),
\]
where $\mathcal{D}(\cdot)$ denotes the DCT operator. By integrating both $L_1$ reconstruction and frequency-domain structure consistency, $L_\mathrm{sa}$ effectively constrains the overall anatomical structure while preserving fine-grained details.

\noindent\textbf{Total Loss} The overall training objective is defined as:
\begin{equation}
	L_\mathrm{total} = 
	  L_\mathrm{seg} 
	+ \lambda_1 L_\mathrm{sc} 
	+ \lambda_2 L_\mathrm{sa}
	+ \lambda_3 L_\mathrm{ldm},
\end{equation}
where $L_\mathrm{seg}$ is a standard hybrid segmentation loss used as auxiliary supervision, $L_\mathrm{ldm}$ is the latent diffusion model denoising loss. 
%
%
%
%
\section{Experiments and Results}

\subsection{Comparisons with State-of-the-Art Methods}

\subsubsection{Datasets and Implementation Details.}
Experiments were conducted on the BraTS2020~\cite{ref_article27,ref_article30} and WMH~\cite{ref_article28} datasets. BraTS2020 contains 369 multi-modal brain MRI scans (T1, T1CE, T2, FLAIR) with expert-annotated tumor segmentation labels, while the WMH dataset includes multiple regions with T1 and FLAIR images as well as white matter hyperintensity (WMH) annotations. All data were preprocessed by axial slicing into $192 \times 192$ 2D images with corresponding masks. The model was implemented in PyTorch 2.1.0 and trained using the Adam optimizer (learning rate $1\times10^{-4}$) with a batch size of 9 on three NVIDIA 4090 GPUs for 100 epochs.

\begin{table}[t]
	\centering
	\small
	\renewcommand{\arraystretch}{1.15}
	\caption{Quantitative comparison on the BraTS 2020 dataset, where $\bar{M}$ denotes the average number of available modalities.}
	\label{tab:quantitative}
	
	\begin{adjustbox}{width=\linewidth}
		\begin{tabular}{c c ccc ccc ccc ccc}
			\hline
			\multirow{2}{*}{$\overline{M}$} 
			& \multirow{2}{*}{Methods}
			& \multicolumn{3}{c}{T1}
			& \multicolumn{3}{c}{T2}
			& \multicolumn{3}{c}{T1CE}
			& \multicolumn{3}{c}{FLAIR} \\
			\cline{3-14}
			
			& 
			& PSNR$\uparrow$ & SSIM\%$\uparrow$ & Dice\%$\uparrow$
			& PSNR$\uparrow$ & SSIM\%$\uparrow$ & Dice\%$\uparrow$
			& PSNR$\uparrow$ & SSIM\%$\uparrow$ & Dice\%$\uparrow$
			& PSNR$\uparrow$ & SSIM\%$\uparrow$ & Dice\%$\uparrow$ \\
			\hline
			
			\multirow{4}{*}{1}
			& MM-GAN 
			& 24.92$\pm$2.31 & 88.75$\pm$1.83 & 0.792
			& 25.10$\pm$2.25 & 89.12$\pm$1.71 & 0.801
			& 26.42$\pm$1.95 & 90.03$\pm$2.12 & 0.809
			& 25.08$\pm$2.14 & 89.65$\pm$1.98 & 0.804 \\
			
			& SynDiff 
			& 25.87$\pm$2.44 & 90.18$\pm$2.65 & 0.821
			& 26.21$\pm$2.58 & 91.02$\pm$3.14 & 0.824
			& 27.38$\pm$2.33 & 92.35$\pm$2.41 & 0.815
			& 26.44$\pm$2.49 & 90.91$\pm$2.87 & 0.818 \\
			
			& MISA-LDM 
			& 27.12$\pm$2.18 & 91.54$\pm$2.22 & 0.852
			& 27.56$\pm$2.36 & 92.03$\pm$2.48 & 0.848
			& 28.74$\pm$2.09 & 93.17$\pm$2.19 & 0.806
			& 27.63$\pm$2.27 & 91.68$\pm$2.41 & 0.829 \\
			
			& \textbf{MSG-LDM} 
			& \textbf{28.05$\pm$2.07} & \textbf{92.48$\pm$2.04} & \textbf{0.871}
			& \textbf{28.33$\pm$2.22} & \textbf{92.67$\pm$2.51} & \textbf{0.868}
			& \textbf{29.41$\pm$2.31} & \textbf{94.12$\pm$1.12} & \textbf{0.821}
			& \textbf{27.92$\pm$2.48} & \textbf{91.95$\pm$2.19} & \textbf{0.842} \\
			
			\hline
			
			\multirow{4}{*}{2}
			& MM-GAN 
			& 25.74$\pm$2.29 & 89.83$\pm$1.70 & 0.812
			& 26.02$\pm$2.20 & 90.35$\pm$1.55 & 0.821
			& 26.91$\pm$1.89 & 91.21$\pm$2.06 & 0.829
			& 26.14$\pm$2.11 & 90.62$\pm$1.91 & 0.823 \\
			
			& SynDiff 
			& 26.68$\pm$2.37 & 91.42$\pm$2.54 & 0.842
			& 27.02$\pm$2.63 & 92.26$\pm$3.08 & 0.845
			& 28.07$\pm$2.42 & 93.36$\pm$2.28 & 0.833
			& 27.15$\pm$2.55 & 91.98$\pm$2.76 & 0.839 \\
			
			& MISA-LDM 
			& 28.01$\pm$2.14 & 92.64$\pm$2.36 & 0.874
			& 28.42$\pm$2.29 & 93.18$\pm$2.49 & 0.869
			& 29.72$\pm$2.05 & 94.12$\pm$2.16 & 0.826
			& 28.58$\pm$2.22 & 92.44$\pm$2.30 & 0.849 \\
			
			& \textbf{MSG-LDM} 
			& \textbf{29.18$\pm$2.04} & \textbf{93.57$\pm$2.10} & \textbf{0.892}
			& \textbf{29.44$\pm$2.18} & \textbf{93.71$\pm$2.63} & \textbf{0.888}
			& \textbf{30.46$\pm$2.28} & \textbf{95.28$\pm$0.95} & \textbf{0.841}
			& \textbf{28.96$\pm$2.61} & \textbf{92.73$\pm$2.18} & \textbf{0.861} \\
			
			\hline
			
			\multirow{4}{*}{3}
			& MM-GAN
			& 27.35$\pm$2.35 & 92.32$\pm$2.88 & 0.861
			& 27.85$\pm$2.87 & 93.18$\pm$3.91 & 0.860
			& 28.65$\pm$2.50 & 94.19$\pm$2.34 & 0.832
			& 27.95$\pm$2.77 & 92.95$\pm$3.13 & 0.856 \\
			
			& SynDiff
			& 28.95$\pm$2.12 & 93.34$\pm$2.41 & 0.896
			& 29.36$\pm$2.32 & 93.95$\pm$2.55 & 0.892
			& 30.65$\pm$2.01 & 94.86$\pm$2.22 & 0.843
			& 29.62$\pm$2.18 & 93.23$\pm$2.35 & 0.868 \\
			
			& MISA-LDM
			& 29.01$\pm$2.02 & 93.86$\pm$2.11 & 0.898
			& 29.66$\pm$2.56 & 94.12$\pm$2.23 & 0.899
			& 30.68$\pm$2.06 & 95.62$\pm$2.12 & 0.849
			& 29.66$\pm$2.18 & 93.28$\pm$2.10 & 0.872 \\
			
			& \textbf{MSG-LDM}
			& \textbf{30.26$\pm$2.03} & \textbf{94.37$\pm$2.02} & \textbf{0.905}
			& \textbf{30.33$\pm$2.27} & \textbf{94.38$\pm$2.89} & \textbf{0.902}
			& \textbf{31.35$\pm$2.40} & \textbf{96.29$\pm$0.80} & \textbf{0.856}
			& \textbf{29.68$\pm$2.74} & \textbf{93.62$\pm$2.11} & \textbf{0.876} \\
			
			\hline
		\end{tabular}
	\end{adjustbox}
\end{table}

\begin{table}[t]
	\caption{Quantitative comparison on the WMH dataset.}
    \centering
    \fontsize{6pt}{7pt}\selectfont
    \renewcommand{\arraystretch}{1.1} 
    \setlength{\tabcolsep}{1pt}      
    \label{tab:quantitative1}
    \begin{tabularx}{\textwidth}{@{} *{7}{>{\centering\arraybackslash}X} @{}}
        \hline
        \multirow{2}{*}{Methods} 
        & \multicolumn{3}{c}{FLAIR $\rightarrow$ T1} 
        & \multicolumn{3}{c}{T1 $\rightarrow$ FLAIR} \\
        \cline{2-7}
        & PSNR$\uparrow$ & SSIM\%$\uparrow$ & Dice\%$\uparrow$
        & PSNR$\uparrow$ & SSIM\%$\uparrow$ & Dice\%$\uparrow$ \\
        \hline
        MM-GAN     & 27.66$\pm$2.86 & 93.68$\pm$2.28 & 0.801
                   & 26.88$\pm$2.79 & 92.78$\pm$2.31 & 0.576 \\
        SynDiff    & 28.42$\pm$2.65 & 94.53$\pm$2.36 & 0.810
                   & 27.89$\pm$2.33 & 93.56$\pm$2.32 & 0.582 \\
        MISA-LDM   & 28.86$\pm$2.25 & 95.23$\pm$2.16 & 0.813
                   & 28.10$\pm$2.13 & 94.65$\pm$2.22 & 0.588 \\
        \textbf{MSG-LDM} & \textbf{29.16$\pm$2.03} & \textbf{96.80$\pm$2.02} & \textbf{0.818}
                   & \textbf{28.38$\pm$2.54} & \textbf{95.55$\pm$2.11} & \textbf{0.595} \\
        \hline
    \end{tabularx}
\end{table}
\subsubsection{Quantitative Results.}
Multi-modal MRI synthesis experiments are conducted by alternately treating each modality as the target and using the remaining modalities together with region masks as conditional inputs. The proposed MSG-LDM is compared with representative multi-modal MRI synthesis methods, including MISA-LDM~\cite{ref_article24}, MM-GAN~\cite{ref_article25}, and SynDiff~\cite{ref_article26}. Experiments are carried out on the BraTS 2020 dataset to evaluate synthesis performance and further validated on the WMH dataset to assess cross-dataset generalization, with quantitative results summarized in Tables~\ref{tab:quantitative} and~\ref{tab:quantitative1}. Evaluation is performed using PSNR, SSIM, and Dice metrics~\cite{ref_article22}, where Dice scores are averaged over the whole tumor (WT), tumor core (TC), and enhancing tumor (ET) regions. Overall, the proposed method consistently outperforms existing approaches across all evaluation metrics.

\begin{figure}[H]
    \centering
    \begin{subfigure}[b]{0.49\textwidth}
        \centering
        \includegraphics[width=1.0\textwidth]{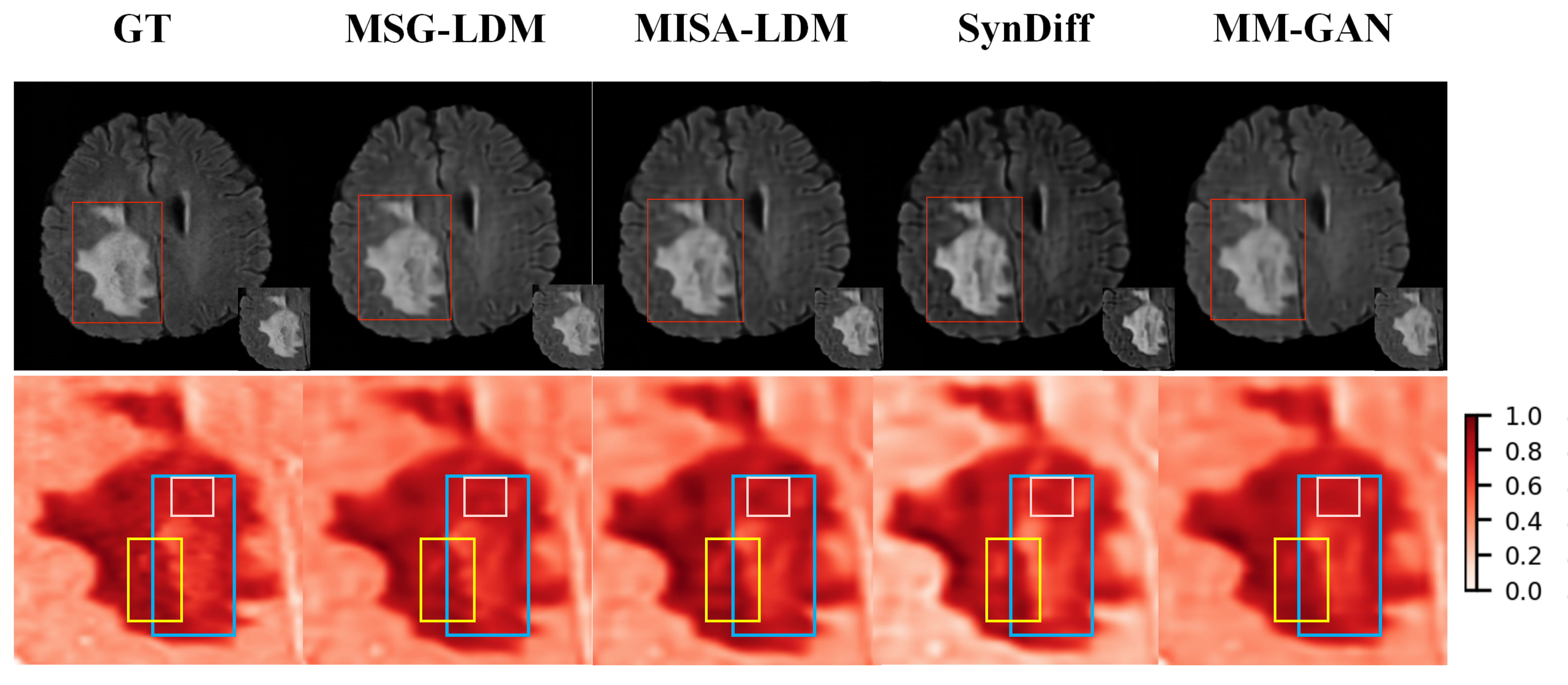}
        \caption{BRATS dataset}
        \label{fig:brats}
    \end{subfigure}
    \hfill
    \begin{subfigure}[b]{0.49\textwidth}
        \centering
        \includegraphics[width=0.99\textwidth]{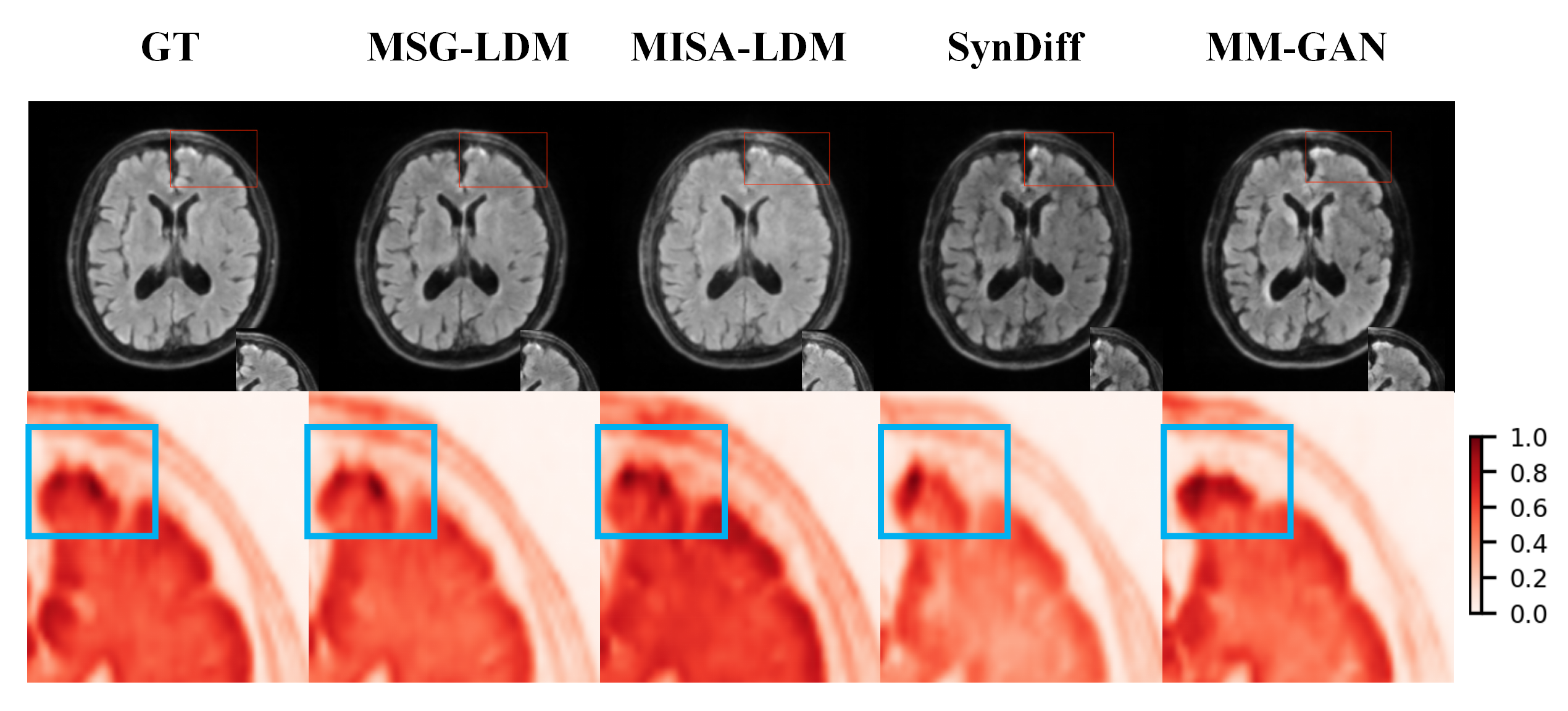}
        \caption{WMH dataset}
        \label{fig:wmh}
    \end{subfigure}
    \caption{Visual comparison of generated images on two datasets.}
    \label{fig:comparison}
\end{figure}
\subsubsection{Qualitative Analysis}
Fig.~3 presents synthesized FLAIR images from the BraTS dataset (a) and the WMH dataset (b). The generated results closely match the ground truth. 
To evaluate structure preservation, we chose the heat maps to examine the structural variation. The results show that the synthesized images by MSG-LDM capture both low-frequency global context and high-frequency fine structural patterns, and exhibits a distribution similar to that of the original images, confirming strong structure consistency.
Fig.~4 shows MRI images synthesized by MSG-LDM under different modality combinations. As the number of available modalities increases, the generated images progressively improve in clarity, structural completeness, and fine-detail representation.
\begin{figure}[H]
	\centering
	\includegraphics[width=0.6\textwidth]{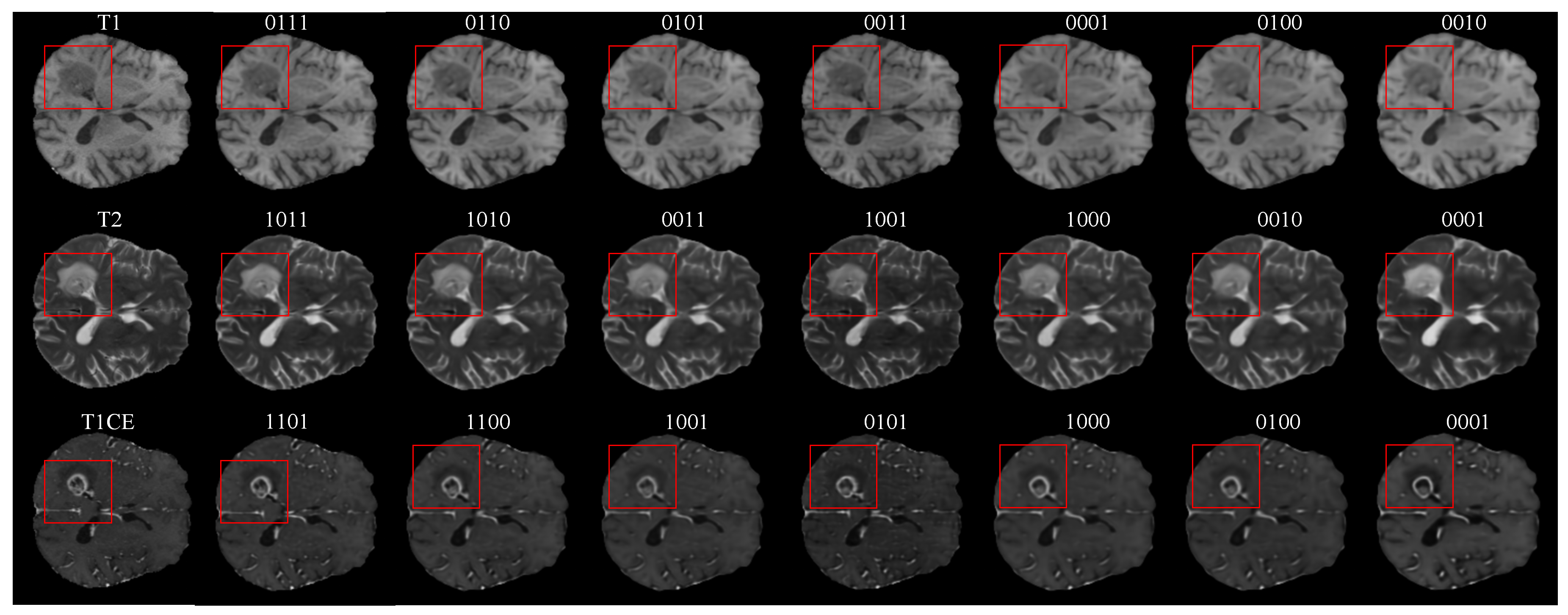}
	\caption{Visual examples of images generated by MSG-LDM. The 4-bit binary code indicate the presence of T1, T2, T1CE, and FLAIR modalities, where “0” means the modality is missing and “1” means it is available.}
	\label{fig1}
\end{figure}

\subsection{Ablation Study }
We mask the FLAIR modality on the BraTS dataset while retaining the remaining three modalities as conditional inputs, and design ablation experiments based on the FLAIR reconstruction task to systematically evaluate the effectiveness of each component. As shown in Table~\ref{tab:ablation_guidance}, removing any module leads to performance degradation in terms of PSNR, SSIM, and Dice, whereas the full model achieves the best results across all evaluation metrics.
\begin{table}[H]
	\centering
	\small
	\caption{Ablation study on the BraTS2020 dataset.}
	\label{tab:ablation_guidance}
	
	\makebox[\linewidth][c]{
		\begin{tabular}{p{4.2cm} p{2cm} p{2cm} p{2cm}}
			\hline
			 & PSNR & SSIM\% & Dice\% \\
			\hline
			w/o Decoupling+MMSF & 27.92 & 92.41 & 85.03 \\
			w/o HFIB                & 28.17 & 92.68 & 85.41 \\
			w/o MSSE                & 29.04 & 93.28 & 86.55 \\
			w/o Loss-sa             & 27.36 & 91.82 & 84.27 \\
			w/o Loss-sc          & 27.11 & 91.54 & 83.89 \\
			Full model (Ours)        & \textbf{29.68} & \textbf{93.62} & \textbf{87.60} \\
			\hline
		\end{tabular}
		
	}
	
\end{table}

\section{Conclusion}
The proposed MSG-LDM explicitly disentangles structure and style information in the latent space, effectively alleviating structural inconsistencies caused by missing modalities in multi-modal MRI translation. By incorporating multi-scale structure modeling together with style consistency and structure-aware constraints, the proposed framework suppresses modality-specific style interference while better preserving key anatomical structures and boundary details, thereby enhancing the stability of structural representations and the reliability of the multi-modal MRI translation. 

\end{document}